\documentclass[aps,twocolumn,showpacs]{revtex4-1}
\usepackage{amsmath,amssymb,graphicx,enumerate}

\newcommand{\nc}{\newcommand}
\nc{\amp}{&}
\nc{\beq}{\begin{equation}}
\nc{\eeq}{\end{equation}}
\nc{\ba}{\begin{eqnarray}}
\nc{\ea}{\end{eqnarray}}
\nc{\alg}{\alpha}
\nc{\algo}{\alpha_{1}}
\nc{\algt}{\alpha_{2}}
\nc{\algi}{\alpha_{i}}
\nc{\algZ}{\alpha_S}
\nc{\algoZ}{\alpha_{1S}}
\nc{\algtZ}{\alpha_{2S}}
\nc{\algiZ}{\alpha_{iS}}
\nc{\mn}{\mu \nu}
\nc{\bet}{\beta}
\nc{\om}{\omega}
\nc{\lom}{\tilde{\omega}}
\nc{\Om}{\Omega}
\nc{\ommin}{\om_{\mbox{\tiny{min}}}}
\nc{\del} {\delta_}
\nc{\dij} {\delta_{ij}}
\nc{\bh} {\bar{h}}
\nc{\gam}{\gamma}
\nc{\Vfar}{V_{\mbox{\tiny{far}}}}
\nc{\Vnear}{V_{\mbox{\tiny{near}}}}
\nc{\Vclass}{V_N}
\nc{\VD}{V_{\mbox{\tiny{mon}}}}
\nc{\love}{k}
\nc{\KK}{J}
\nc{\lr}{\tilde{r}}
\nc{\pref}{\zeta}
\def\vx{\vec x}
\def\vy{\vec y}

\begin{document}

\title{Quantum Gravitational Force Between Polarizable Objects}

\begin{abstract}
Since general relativity is a consistent low energy effective field theory, it is possible to compute quantum corrections to classical forces. Here we compute a quantum correction to the gravitational potential between a pair of polarizable objects. We study two distant bodies and compute a quantum force from their induced quadrupole moments due to  two graviton exchange. The effect is in close analogy to the Casimir-Polder and London-van der Waals forces between a pair of atoms from their induced dipole moments due to two photon exchange. The new effect is computed from the shift in vacuum energy of metric fluctuations due to the polarizability of the objects. We compute the potential energy at arbitrary distances compared to the wavelengths in the system, including the far and near regimes. In the far distance, or retarded, regime, the potential energy takes on a particularly simple form: $V(r)=-3987\,\hbar\,c\,G^2\algoZ\,\algtZ/(4\,\pi\,r^{11})$, where $\algoZ,\,\algtZ$ are the static gravitational quadrupole polarizabilities of each object. We provide estimates of this effect.
\end{abstract}

\pacs{04.60.Bc, 03.70.+k, 04.30.-w, 42.50.Lc}

\author{L.~H.~Ford$^*$, Mark P.~Hertzberg$^\dagger$, J.~Karouby$^\ddagger$}
\affiliation{Institute of Cosmology, Department of Physics and Astronomy\\
Tufts University, Medford, Massachusetts 02155, USA}

\date{\today}

\maketitle

{\em Introduction}.---Quantum gravitational effects are often thought to depend on the details of the unknown full theory of quantum gravity. This is true for phenomena near the Planck energy. But at low energies, general relativity is a consistent effective field theory, so this allows, in principle, access to low energy quantum effects without knowledge of the microscopic details. For example, this reasoning led to the determination of a quantum correction to the Newtonian potential by Donoghue \cite{Donoghue} and others \cite{EFTOthers}, with result $\VD(r)=-41\,\hbar\,G^2 M_1M_2/(10\,\pi\, c^3\,r^3)$. This particular effect applies to particles, from summing one-loop Feynman diagrams with off-shell gravitons. It is a quantum force between point-masses, or {\em monopoles}. Corrections from induced higher order multipoles have not, to our knowledge, been computed in the literature. However, realistic astrophysical structures, such as stars, carry non--zero quadrupole polarizability, so higher order multipoles can contribute.

\begin{figure}[t]
\includegraphics[width=\columnwidth]{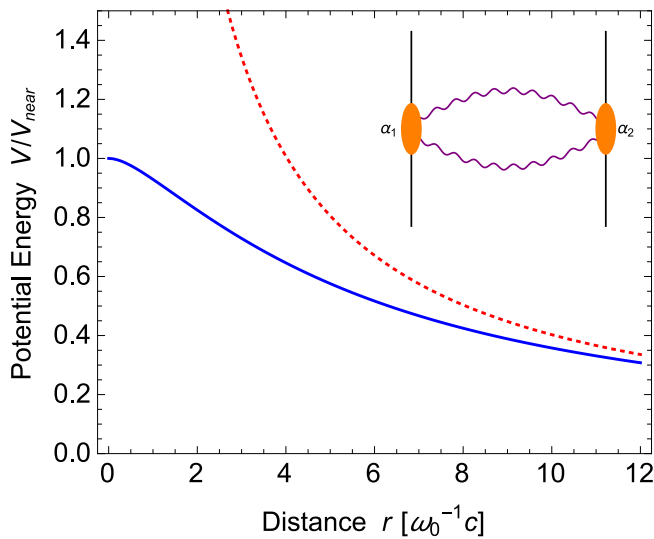}
\caption{Potential energy $V$ in simple harmonic model divided by near result $\Vnear$ as a function of distance $r$ (in units $\om_0^{-1}c$, with $\om_{1}=\om_{2}=\om_0$). Solid (blue) is the exact result $V(r)/\Vnear(r)$. Dotted (red) is the far result $\Vfar(r)/\Vnear(r)$. Inset is corresponding loop diagram of gravitons (purple wiggly lines) between polarizable sources (orange disks).}
\label{Vfigure}
\end{figure}

In this Letter we shall rigorously compute a low energy quantum gravitational correction to the force between any pair of localized polarizable objects. We determine a force from their induced gravitational {\em quadrupole} moments due to quantum fluctuations in the metric. Diagrammatically, the effect is associated with two graviton exchange from 4-point vertices between the sources; see inset in Fig.~\ref{Vfigure}. To determine this, we shall use the method of Casimir \cite{Casimir}, whereby we compute a shift in the vacuum energy due to the polarizability of the objects.

In the electromagnetic case, quantum induced dipole moments lead to the Casimir-Polder force between atoms or molecules in the far, or retarded, regime \cite{CasimirPolder} and to the London-van der Waals force in the near, or instantaneous, regime \cite{London}. The potential energy is found to scale as the product of the electric dipole polarizabilities and is $\propto 1/r^7$ in the retarded regime and $\propto 1/r^6$ in the instantaneous regime, where $r$ is the distance between the two bodies. For reviews see \cite{CasimirReview}. Various aspects of these quantum electromagnetic forces have been studied, including applications \cite{Concrete}, experimental verification \cite{Experiment}, shape dependence \cite{Shape}, cosmology \cite{CurvedSpacetime}, etc.

Since metric fluctuations are sourced by quadrupole moments, rather than dipole moments, the leading gravitational effects carry a different scaling with distance $r$. In this Letter we derive the following quantum gravitational contribution to the potential energy in the ``far" and ``near" regimes
\ba
\Vfar(r)\amp=\amp-\frac{3987\,\hbar\,c\,G^2 }{4\pi\,r^{11}} \algoZ\,\algtZ, \label{far} \\
\Vnear(r)\amp=\amp-\frac{315\, \hbar\, G^2 }{ \pi\,r^{10}} \!\int_0^\infty d\om \,\algo(i\,\om)\,\algt(i\,\om),\, \label{near}
\ea
where $\algi(\om)$ (with $\algiZ\equiv\algi(0))$ are the gravitational quadrupole polarizabilities, to be discussed later. We also determine the full dependence at intermediate distances.

As opposed to the monopole-monopole quantum effect mentioned at the beginning, this quadrupole-quadrupole quantum effect involves on-shell metric fluctuations, but whose amplitude is set by quantum mechanics. 

So, in this Letter, we begin by solving for the Riemann tensor in classical general relativity due to a quadrupole moment with frequency $\om$. Then we introduce the gravitational polarizability and determine the conditions for eigenfrequencies of a two body system. Next we compute the quantum vacuum energy from a sum over these eigenfrequencies. Finally we determine the force between the two objects and study it in specific cases.

{\em Riemann Tensor From Quadrupole}.---We begin by expanding the metric around flat space as $g_{\mu\nu}=\eta_{\mu\nu}+h_{\mu\nu}$ and define the trace-reversed perturbations $\bar{h}_{\mu\nu}=h_{\mu\nu}-\frac{1}{2} \eta_{\mu\nu} h$,  with $h\equiv h_\mu^\mu ,\ \bar{h}=-h$. Then working in the linearized approximation in the Lorenz gauge ($\partial_\mu \bar{h}^{\mu\nu}=0$), the Einstein equations become $\Box \bh_{\mu \nu }= -16 \pi G T_{\mn}$ (for convenience we work in units $c=1$ and re-instate $c$ in our final results). The only contribution to $\bar{h}_{\mu\nu}$ that will be associated with a force, arises from the source $T_{\mu\nu}$. So we ignore any solutions of the homogeneous equation $\Box \bh_{\mu\nu}=0$ and focus on the particular solution
\beq
\label{hdef}
\bar{h}_{\mn}(t,\vx)=4G\! \int \! d^3y\, \frac{T_{\mn}(t-|\vx-\vy |,\vy)}{|\vx-\vy |},
\eeq
where the energy-momentum tensor is evaluated at the retarded time $t-|\vx-\vy|$.

Assuming we have a localized source, $R\ll r$ where $R$ is the typical size of the source and $r\equiv|\vec{x}|$, we Taylor expand the retarded energy-momentum tensor to second order in $|\vec{y}|/r\ll 1$. By Fourier transforming in time and carrying out the spatial integral $\int\!d^3y$ using local conservation $\partial_\mu T^{\mu\nu}=0$, we find the following expressions for the $00$, $0i$, and $ij$ components of the metric
\ba
\label{h1}
\bar{h}_{00}(\om,\vec{x})\amp=\amp2Ge^{i\om r}\Big{[}-\frac{\om^{2}}{r^{3}}x^{i}x^{j}I_{ij}(\om)\nonumber\\
&&\,\,\,\,\,\,\,\,\,\,\,\,\,\,\,\,+3\left(\frac{1}{r^{5}}-i\frac{\om}{r^{4}}\right)x^{i}x^{j}Q_{ij}(\om)\Big{]}, \\
\bh_{0i}(\om ,\vec{x} ) \amp=\amp2G e^{i\om r} \! \left(\frac{i}{r^3}+\frac{\om}{r^2}\right)  \om\, x^j I_{ij}(\om),\\
\bh_{ij}(\om ,\vec{x} ) \amp=\amp-2G e^{i\om r}{\om^2\over r} I_{ij}(\om),
\ea
where $I_{ij}$ is the quadrupole moment and $Q_{ij}$ is its traceless part
\ba
I_{ij}(\omega) \amp\equiv\amp\int \!d^3y\, \rho(\omega,\vec{y})\, y_i\, y_j,\\
Q_{ij}(\omega)\amp\equiv\amp I_{ij}(\omega)-\frac{1}{3}I(\omega)\,\delta_{ij}.
\ea
Here $\rho\equiv T^{00}$ is the energy density and $I\equiv \delta^{ij}I_{ij}$. We have taken $\om>0$ which ignores the monopole.

We note that when studying time varying quadrupoles in the literature, one often tracks the terms $\propto 1/r$ only, which dominate at large distances $r\gg \omega^{-1}$. However, here we are interested in regimes where $r$ may be smaller or larger than $\omega^{-1}$, so we have computed the complete expression to quadrupole order. Our only assumption is that the size of the source is small compared to other scales in the problem: $R\ll r$ and $R\ll \omega^{-1}$.

As we will explain, the needed spatial components of the Riemann tensor are $R_{0i0j}$. A straightforward, but technical calculation using the above $\bh_{\mn}$, gives the following result in this linearized approximation
 \beq
 \label{RQ}
 R_{0i0j}(\omega,\vec{x})=\frac{G e^{ir\omega}}{r^5}\Lambda_{ij}^{kl}(\om\,r,\vec{n})\,Q_{kl}(\om),
 \eeq
where $\Lambda^{kl}_{ij}$ is a dimensionless 4-index projection tensor that depends on the dimensionless product $\gam\equiv \om\,r$ and unit vector $\vec{n}\equiv \vec{x}/r$. We find $\Lambda_{ij}^{kl}$ to be
\ba
&&\Lambda_{ij}^{kl}(\gam,\vec{n})=-{1\over2}\Big{[}(6-6\gam^2+2\gam^4-6i\gam+4i\gam^3)\delta_i^k\delta_j^l\nonumber\\
&&\,\,\,\,\,\,+(-15+9\gam^2-\gam^4+15i\gam-4i\gam^3)(n^i n^k\delta_j^l+\mbox{perm})\nonumber\\
&&\,\,\,\,\,\,+(-15+3\gam^2+\gam^4+15i\gam+2i\gam^3)\delta_{ij}n^kn^l\nonumber\\
&&\,\,\,\,\,\,+(105-45\gam^2+\gam^4-105i\gam+10i\gam^3)n^in^jn^kn^l\Big{]},
\ea
where ``perm" represents 3 more terms from permuting $i\leftrightarrow j$ and $k\leftrightarrow l$.

{\em Gravitational Polarizability}.---The linearized Einstein equations may be written as Maxwell-like equations, yielding the gravito-electromagnetic equations. In the non-relativistic regime, the gravito-electric field is related to the Riemann tensor by $E_{ij}=R_{0i0j}$. We can define a gravitational quadrupole polarizability $\alg$ as the constant of proportionality between an {\em applied} gravito-electric field $R_{0i0j}$ and an {\em induced} quadrupole moment $Q_{ij}$, as follows
\ba
Q_{ij}(\omega)= \alg(\omega)\, R_{0i0j}(\omega,\vec{x}),
 \label{QaR}
\ea
where $\vec{x}$ is evaluated at the location of the object it is polarizing. For simplicity, we are assuming the system is isotropic, so the polarizability $\alg$ does not carry any tensor indices. This definition is analogous to that of isotropic dipole polarizability in electromagnetism. In general $\alg$ will depend on frequency. Dimensionally, we have $[\alg] = L^5/G$.

{\em Eigenstate Equation}.---Consider two polarizable objects distant from each other, one located at $\vec{x}_1$ with quadrupole polarizability $\algo$, the second at $\vec{x}_2$ with $\algt$. A time dependent quadrupole on the first object $Q^1_{ij}$ radiates a gravitational field $R^1_{0i0j}$ which  polarizes the second object. The second object then acquires a time dependent quadrupole $Q^2_{ij}$ which in turn radiates $R^2_{0i0j}$ and polarizes the first object. In order for this cycle to be in phase, the system must organize into normal modes. Using Eqs.~(\ref{RQ},\,\ref{QaR}), this implies the following conditions must be satisfied
\ba
&&Q^1_{ij}(\om)={G e^{ir\omega}\over r^5}\algo(\om) \,\Lambda_{ij}^{kl}(r\,\om,\vec{n}_1)\,Q^2_{kl}(\om), \\
&&Q^2_{ij}(\om)={G e^{ir\omega}\over r^5}\algt(\om) \,\Lambda_{ij}^{kl}(r\,\om,\vec{n}_2)\,Q^1_{kl}(\om).
\ea
 Combining this pair of equations gives the eigenstate equation
 \beq
 M_{ij}\equiv K_{ij}^{kl}(\om,r,\vec{n}_1,\vec{n}_2)\,Q_{kl}(\om)=0,
 \eeq
 where
  \ba
 &&K_{ij}^{kl}(\om,r,\vec{n}_1,\vec{n}_2)= \delta_i^k\delta_j^l \nonumber\\
 && - \frac{G^2 e^{2 i r\omega}}{r^{10}}\algo(\om)\,\algt(\om)\, \Lambda_{ij}^{mn}(r\,\om,\vec{n}_1)\,\Lambda_{mn}^{kl}(r\,\om,\vec{n}_2).
 \,\,\,\,\,\,\,\,
\ea

Without loss of generality we choose to place the two sources separated along the $x$-axis, by setting $\vec{n}_1=(1,0,0)$ and $\vec{n}_2=(-1,0,0)$. By evaluating the above matrix $M_{ij}$, we obtain
 \beq
 \label{eigen}
M_{ij}=
\begin{pmatrix} 
 Q_{11} F_1 & Q_{12} F_2 & Q_{13} F_2 \\
 Q_{12} F_2 & Q_{11} F_-+Q_{22}F_3 & Q_{23} F_3 \\
 Q_{13} F_2 & Q_{23} F_3 &  -Q_{11} F_+-Q_{22} F_3
 \end{pmatrix}\!,\,\,\,\,\,\,\,
 \eeq
 where $F_\pm\equiv (F_3\pm F_1)/2$. Here we have suppressed the arguments $M_{ij}=M_{ij}(\om,r)$, $Q_{ij}=Q_{ij}(\om)$, and $F_a=F_a(\om,r)$ ($a=1,2,3$). The $F_{a}$ are found to be
 \ba
F_1(\om, r)&=& 1 - 36\KK(\om,r) [3 - 3i\gam-\gam^2]^2,  \label{F1}\\
F_2(\om, r)&=& 1 - 4 \KK(\om,r) [6-6i\gam-3\gam^2+i\gam^3]^2,  \label{F2}\\
F_3(\om, r)&=& 1 - \KK(\om,r) [3-3i\gam-3\gam^2+2i\gam^3+\gam^4]^2,  \label{F3}\,\,\,\,\,\,\,\,\,
 \ea
 ($\gam\equiv \om\,r$), where
 \beq
\KK(\om,r)\equiv\frac{G^2 e^{2i\om r}}{r^{10}} \algo(\om)\, \algt(\om).
\label{KKdef} \eeq

{\em Eigenstates}.---The matrix equation $M_{ij}(\om,r)=0$ is only satisfied by special frequencies and quadrupole moments. This defines a set of eigenstates. Since we have quadrupole moments, we have eigenmatrices $Q_{ij}(\om)$ (instead of eigenvectors $p_i(\om)$ that appear in the electromagnetic case \cite{Method}). By searching for solutions of $M_{ij}=0$, we are able to systematically determine each of the eigenstates, which we find come in 3 distinct cases:
\begin{enumerate}[{(a)}]
\item A set of non-degenerate eigenstates
\beq
Q_{ij}^{(a)}=
\begin{pmatrix}
Q_{11}&0& 0\\
0 & \frac{-Q_{11}}{2}&0\\
0 &0 &\frac{-Q_{11}}{2}
\end{pmatrix}
\,\,\,\,\mbox{with}\,\,\,\, F_1(\om,r)=0.
\eeq
\item A set of doubly degenerate eigenstates
\beq
Q_{ij}^{(b)}=
 \begin{pmatrix}
0& Q_{12}&Q_{13}\\
Q_{12} & 0&0\\
Q_{13} &0 &0
\end{pmatrix}
\,\,\,\,\mbox{with}\,\,\,\, F_2(\om,r)=0.
\eeq
\item A set of doubly degenerate eigenstates
\beq
Q_{ij}^{(c)}=
  \begin{pmatrix}
0&0& 0\\
0 & Q_{22}&Q_{23}\\
0 &Q_{23} &-Q_{22}
\end{pmatrix}
\,\,\,\,\mbox{with}\,\,\,\, F_3(\om,r)=0.
\eeq
\end{enumerate}
This provides the complete set of eigenstates of the system. By solving the equations $F_1=F_2=F_3=0$ the eigenfrequencies are specified.

{\em Vacuum Energy}.---The leading quantum correction to the potential energy arises from a one-loop effect from two graviton exchange; see inset in Fig.~\ref{Vfigure}. This may also be viewed as a shift in vacuum energy due to the polarizability of the objects. The corresponding potential energy is given by the sum over eigenfrequencies $V(r) = \sum_n\hbar\,\om_n/2-\sum_n\hbar\,\lom_n/2$. We have indicated that $V=V(r)$ since the eigenfrequencies depend on the distance between the sources.  The frequencies $\om_n$ refer to the system at some finite distance $r$. The frequencies $\lom_n$ refer to the system at some large separation $\lr$; we shall take the limit $\lr\to\infty$ at the end of the calculation. We subtract off the latter as only energy differences are important.

Following a technique used in the electromagnetic case \cite{Method}, we make use of {\em Cauchy's argument principle} \cite{ArgumentPrinciple} in order to convert these sums to integrals. The argument principle states that for an analytic function $g(z)$ and a meromorphic function $f(z)$
\beq 
{1\over 2\pi i}\oint_{C} dz\,g(z)\, \partial_z \ln[f(z)] = \sum_n g(z_n) - \sum_p g(z_p), 
\label{Cauchy}\eeq
where $C$ is a simple contour. The first sum is over all zeros $z_n$ of $f(z)$ counted with their degeneracies and the second sum is over all poles $z_p$ of $f(z)$ counted with their orders.

We choose $g(z)=\hbar\,z/2$ and $f(z)=F(z,r)/F(z,\lr)$,
where the ``master" function $F(\om,r)$ that we need to integrate has zeroes for each eigenfrequency of the system. It is given by the product
\ba
 F(\om ,r)= F_1(\om ,r)\,F_2(\om ,r)^2\,F_3(\om ,r)^2,
\label{Fproduct} \ea
where the various powers come from the degeneracy for each corresponding eigenfrequency: (a), (b), (c). 

We note that the poles of $F(z,r)$ are identical to the poles of $F(z,\lr)$, since they can only arise from poles of $\alpha_i(z)$, which are independent of $r$. So the poles of $F$ cancel out of $f$. The argument principle then says that the vacuum energy $V(r) = \sum_n\hbar\,\om_n/2-\sum_n\hbar\,\lom_n/2$ can be expressed as
\beq
V(r)= \frac{1}{2 \pi i} \oint_C dz\, \frac{\hbar\, z}{2}\, \partial_z \ln\!\left[F(z,r)\over F(z,\lr)\right],
\eeq 
 where the contour encloses all the zeros of $F$. Deforming the contour to an integral over the positive imaginary axis and integrating by parts gives
 \beq
V(r)= \frac{\hbar}{2 \pi } \int_0^\infty d\om\, \ln\!\left[F(i\,\om,r)\over F(i\,\om,\lr)\right],
\label{Vimag}\eeq
with $z=i\,\om$.

{\em Quantum Gravitational Potential}.---Substituting the expression for $F$ in Eqs.~(\ref{F1}--\ref{F3},\,\ref{KKdef},\,\ref{Fproduct}) into Eq.~(\ref{Vimag}), taking $\lr\to\infty$, and Taylor expanding the logarithm to leading order for large $r\gg R$ (as our analysis is only valid in that regime) we obtain our primary result (re-instating factors of $c$)
\beq
V(r)= -\frac{\hbar\, G^2}{ \pi\,r^{10} } \int_0^\infty d\om\,S(\om\,r/c)\,
\algo(i\,\om)\,\algt(i\,\om),
\label{main}\eeq
where
\ba
&&S(\gam) \equiv e^{-2\gam}
\big{[}315+630 \gam+585 \gam^2+330 \gam^3\nonumber\\
&&\hspace{1.8cm}+129 \gam^4+42 \gam^5+14 \gam^6+4 \gam^7+\gam^8\big{]} 
\label{Sdef}\ea
$(\gam\equiv\om\,r/c$). 
For any given choice of gravitational polarizabilities $\algi(\om)$ ($i=1,2$) this integral can, in principle, be performed, giving a definite quantum gravitational correction to the force between polarizable objects.

{\em Limiting Cases}.---In the far regime $r\gg \om_0^{-1}c$, where $\om_0$ is some characteristic frequency of response in the system, the exponential in $S$ enforces that only the static polarizability $\algi(i\,\om)\to\algiZ\equiv\algi(0)$ contributes. Then carrying out the integral, we obtain the result reported earlier in Eq.~(\ref{far}). This is analogous to the Casimir-Polder force between electrically polarizable objects \cite{CasimirPolder}.

In the near regime $r\ll \om_0^{-1}c$, we can replace $S(\om\,r/c)\to S(0)=315$, to simplify $V$. This gives the result reported earlier in Eq.~(\ref{near}). Note that factors of $c$ cancel out in this regime, as this is the non-relativistic limit. This is analogous to the London-van der Waals force between electrically polarizable objects \cite{London}.

{\em Simple Harmonic Model}.---For realistic systems, the gravitational polarizability $\algi(\om)$ of each object can be a complicated function of frequency. However, for definiteness, let us consider a simple model in which each object's response is characterized by a single harmonic oscillator with frequency $\om_{i}$. The polarizabilities may be written as \cite{HarmonicPolarizability}
\beq
\algi(\om)={\algiZ\over 1-(\om/\om_{i})^2}.
\eeq

In this case the integral over frequency in the near regime may be performed analytically, with the result
\beq
\Vnear(r)=-\frac{315\, \hbar\, G^2 }{2\,r^{10}} {\om_{1}\,\om_{2}\over\om_{1}+\om_{2}}\,\algoZ\,\algtZ.
\label{near2}\eeq
Furthermore the full result in Eqs.~(\ref{main},\,\ref{Sdef}) can be determined. Although we do not have an analytical expression for the full integral, the result can be readily determined numerically. In Fig.~\ref{Vfigure} we plot the potential $V(r)$ divided by the near approximation $\Vnear(r)$ as a function of distance $r$. We also include the far approximation $\Vfar(r)$ for comparison.

For a simple system, such as an elastic sphere, one can compute the static polarizability $\algiZ$ in this simple harmonic model. One should apply the geodesic deviation equation $D^2\!\Delta^\mu/d\tau^2=R^\mu_{\,\,\,\nu\rho\sigma}U^\nu U^\rho\Delta^\sigma$, in the non-relativistic limit, to each infinitesimal part of the system and integrate, using Newton's law for each infinitesimal spring. If the object has radius $R_i$, mass $M_i$, frequency $\om_i$, then dimensional analysis readily gives $\algiZ\sim M_iR_i^2/\om_i^2$. Now an important reference frequency is $\Om_i\equiv\sqrt{G M_i/R_i^3}$ which is of the order of the orbital frequency of a gravitationally bound system. One expects $\Om_i$ to set a lower bound on $\om_i$ for any physical system. 

Altogether we can approximate the far and near regimes as
\ba
\Vfar(r)\amp\sim\amp-
{\hbar\,c\over r}\left(\Om_1\,\Om_2\over\om_1\,\om_2\right)^{\!2}{R_1^5\,R_2^5\over r^{10}},\label{farb}\\
\Vnear(r)\amp\sim\amp-\hbar\,\ommin\left(\Om_1\,\Om_2\over\om_1\,\om_2\right)^{\!2}{R_1^5\,R_2^5\over r^{10}}.\label{nearb}
\ea
($\ommin\equiv\mbox{Min}\{\om_1,\om_2\}$).
For electrically bound systems, such as atoms, $\om_i\gg\Om_i$, the result is suppressed. For gravitationally bound systems, $\om_i\sim\Om_i$, the result is largest. 

In the near regime, we compare to the monopole-monopole result \cite{Donoghue,EFTOthers} $\VD(r)\sim-\,\hbar\,G^2 M_1M_2/(c^3\,r^3)$. We get $\Vnear(r)/\VD(r)=\mathcal{O}((\Om/\om_0)^4(R/R_S)^{3/2}(R/r)^7)$, where $R_S$ is the Schwarzschild radius. The regime of validity of our quadrupole-quadrupole result is formally $R\ll r$, but taking this hierarchy of scales to be modest and $\om_0\sim\Om$, this new result can in principle be larger than the monopole-monopole correction for $R\gg R_S$.

{\em Stars}.---For a more careful analysis of $\alg$ in realistic gravitationally bound compact systems, we turn to the literature, where it has been computed for various kinds of astrophysical objects \cite{GravPolarOther}. Dimensional analysis $[\alg]=L^5/G$ is suggestive of the scaling. To extract a dimensionless measure of polarizability, one normally defines the so-called ``Love number" $\love \equiv -3\,G\,\algZ/(2 R^5)$. The Love number depends primarily on the object's compactness $C\equiv GM/(R\,c^2)$ and equation of state. In the case of neutron stars and self-bound quark stars, for example, \cite{GravPolarPostnikov} finds values of order $\love=\mathcal{O}(10^{-1})$, depending on parameters. In the far regime (applicable for distant binaries) we have $\Vfar(r)=-\beta\,\hbar\,c\,R_1^5\,R_2^5/r^{11}$, with $\beta\equiv 443\,\love_1\,\love_2/\pi=\mathcal{O}(1)$. Consistency requires $R_{1,2}\ll r$, so this gives $|\Vfar(r)|\ll\hbar\,c/r$; an extremely small energy compared to the classical Newtonian potential energy $\Vclass(r)=-G\, M_1 M_2/r$.

{\em Black Holes}.---A natural object for the study of quantum gravity effects is a black hole. Some authors have claimed that the Love number of Schwarzschild black holes is $\love=0$ \cite{BlackHoles}. This would imply our quantum force vanishes in this case, though it may be non-zero for other types of black holes.

{\em Discussion}.---We have computed a new quantum gravitational force in the low energy regime between any pair of localized objects. This complements the point-particle result in the literature \cite{Donoghue,EFTOthers}. This is a rare, precise, and definite prediction of quantum gravity, independent of the details of its UV completion. So a consistency test of any candidate UV complete theory is that this force is recovered at low energies. The general result is given in Eqs.~(\ref{main},\,\ref{Sdef}), with far and near limits given in Eqs.~(\ref{far},\,\ref{near}). Some given applications are to simple harmonic models, neutron stars and black holes.

Unlike the point-particle result, this new force depends on the intrinsic material properties of the objects through their gravitational quadrupole polarizabilities (analogous to the electromagnetic Casimir-Polder and London-van der Waals forces). As such it represents a type of quantum violation of the equivalence principle. A clear example of equivalence principle violation is in Ref.~\cite{EquivPrinciple}, whereby the motion is different between spin 1 and spin 0 particles. In contrast here we see a difference in motion due to different gravitational polarizabilities.

Of course in ordinary circumstances, the effect is extremely small. In order to increase the size of this quantum force, relative to the classical force, one needs small, nearby objects and negligible corrections from other forces. A speculative example of this sort, would be nearby microscopic clumps built out of heavy sterile neutrinos.

Future work includes extending these quadrupole-quadrupole quantum gravity results to other multipole moment contributions. Potentially there can be monopole-quadrupole cross terms, and others, that could be interesting. The monopole-monopole, or point-particle, result for the potential energy is $\propto 1/r^3$, while our new quadrupole-quadrupole result, in the retarded regime, is $\propto 1/r^{11}$. We expect a monopole-quadrupole contribution to be $\propto 1/r^7$. 

Our calculations use the quantum vacuum energy of metric fluctuations in the presence of polarizable sources, while the point-particle calculations utilize a sum over Feynman diagrams of quantum particles. Future work would be to establish a unified framework to compute all such phenomena, perhaps through the development of effective Feynman vertices for polarizable sources.

{\em Acknowledgments}.---We would like to thank Audrey Mithani, Ken Olum, Leo Stein, and Alexander Wiegand for discussion. We also acknowledge the Institute of Cosmology at Tufts University for support. LHF acknowledges support from NSF Grant PHY-1506066.

{\em Email Correspondence.--- $^*$ford@cosmos.phy.tufts.edu, 
$^\dagger$mark.hertzberg@tufts.edu, $^\ddagger$johanna.karouby@tufts.edu}

\bibliographystyle{plain}

\end{document}